# Double adiabatic theory of collisionless geodesic acoustic modes in tokamaks

A. B. Hassam and R. G. Kleva
*University of Maryland, College Park*

**Abstract**

Collisionless geodesic acoustic modes in tokamaks being supersonic for large "safety factor" $q$, the CGL (G. Chew, M. Goldberger, F. Low, 1956)[1] double-adiabatic fluid closure is applied to formulate a theory for these modes. The basic linear normal mode is obtained. External means to drive these modes at resonance, as has been proposed earlier, are explored. The external drivers considered include external magnetic forces to effect flux surface displacements, as well as non-axisymmetric ion heating. Finally, the damping of these modes from collisional magnetic pumping is investigated using a model set of CGL collision-corrected equations.

**Introduction**

The geodesic acoustic mode (GAM) is an axisymmetric oscillation of a toroidal magnetically confined plasma, resulting from an interplay between poloidal plasma rotation and the perpendicular flux tube compression that ensues as the flux tube moves between differing B field strength regions poloidally.[2] GAMs have been observed on several tokamaks;[3,4] they are thought to play an important role, given the zonal nature of the associated flows,[5-7] in determining turbulence saturation levels in tokamaks.[3-7]

The theory of GAMs is well developed. Both collisional fluid theories[2,8] and collisionless kinetic theories,[5-7,9-12] have been presented. Damping from Landau effects in the collisionless limit has also been calculated. For the present paper, we note that since the frequency of this mode, $\omega_{GAM}$, is super-parallel-sonic, ie, $\omega_{GAM} \sim$ *(ion thermal speed)/R* is greater than the parallel acoustic mode frequency by a factor of the tokamak "safety factor" $q$, collisionless geodesic acoustic modes in tokamaks can be described by the CGL (G. Chew, M. Goldberger, F. Low, 1956) double-adiabatic fluid closure[1] for the ions. This allows, at least to lowest order, a somewhat simpler and nonlinear formulation. In this paper, we develop these equations to study collisionless GAMs in tokamaks. We show that the linear normal mode frequency from CGL theory agrees with that found by Sugama and Watanabe[11] and others,[12] from the full kinetic theory.

A further motivation for the present study stems from recent proposals by Hallatschek and McKee[13], Cowley[14], and by McKee and others[15] to drive GAMs on the D3D tokamak at resonance using the I-coil system or by ICRF heating.[16] We add external drivers in our CGL theory as a simple extension to allow a study of resonantly driven GAMs. The drivers in our theory include external magnetic forces to effect flux surface displacements as well as sources to provide modulated non-axisymmetric ion heating.

Finally, since the GAM oscillation involves motion of toroidal flux tubes back and forth between regions of differing B, energy is "pumped" into $p_\perp$ during a compression resulting in damping from "magnetic pumping, as is well-known.[17-21] This damping would be in addition to any collisionless Landau damping and could, in some cases, be larger than the latter. We calculate a damping rate from magnetic pumping using a set of model collision-corrected CGL equations introduced earlier.[20,21]

In the next section, we present the basic equations in the collisionless limit and perform a linearization to calculate a frequency for GAMs. In Section 3, we add external drivers to this calculation to investigate resonantly driven GAMs, an experimental idea suggested by Hallatschek and McKee and by Cowley. In Sec 4, we define the collision-corrected CGL equations and reconsider the linear mode to extract a damping rate for magnetic pumping.

**Equations**

GAMs fall in the ideal magnetohydrodynamic (MHD) regime of plasma description in that the frequency of the mode is in the sonic range, the wavelengths are macroscopic (much longer than the ion gyroradius), and thus the frequencies are super-diamagnetic and the flows are dominated by ExB and parallel flows. As a consequence, one may apply the ideal MHD ordering and use the ideal MHD equations. However, in the collisionless limit, the pressure is tensorial, given by

$$\vec{\vec{P}} = p_\perp (\vec{\vec{1}} - \hat{b}\hat{b}) + p_\parallel \hat{b}\hat{b} \qquad (1)$$

where $\hat{b} = \vec{B}/B$. As shown by Kulsrud,[22] the appropriate collisionless MHD system of equations is akin to the usual ideal MHD system, governed by the equation of continuity and the momentum equation with tensorial pressure; however, the parallel and perpendicular pressure equations of state need to be computed from the drift-kinetic equations (DKE) in the MHD ordering, *viz.,*

$$\partial f / \partial t + v_\parallel \nabla_\parallel f + \vec{u}_\perp \cdot \vec{\nabla} f + (-\mu \nabla_\parallel B + eE_\parallel / m) \partial f / \partial v_\parallel = 0. \qquad (2)$$

Here, the DKE is to be applied to each species, with the species distribution function being $f = f(v_\parallel, \mu, \vec{x}, t)$; $\mu = v_\perp^2 / (2B)$, $\vec{u}_\perp$ is the ExB drift, and $E_\parallel$ is the parallel electric field. The quasineutrality condition determines $E_\parallel$. For long wavelength GAMs, $E_\parallel$ is small (as we will elaborate on later).

To simplify the description, two other usual approximations may be made.[2,5-12] First, for low $\beta$, it may be assumed that the magnetic field is approximately static and the response is electrostatic; thus $\vec{E} \approx -\vec{\nabla}\varphi$. In addition, since the frequency is ion-sonic, the electron response is adiabatic and, thus, the parallel electric field is small, yielding $\varphi \approx \varphi(\psi)$, where $\psi$ is the magnetic flux surface label. The remaining governing equations can be written as a system of "magnetic differential equations" for the variables $\{n, u_\parallel, \varphi\}$, with the pressure tensor elements $\{p_\perp, p_\parallel\}$ determined by the DKE. A second approximation may also be made: the GAM frequency scales as $v_{thi}/R$, whereas the bounce frequency, $v_{thi}/qR$, is much lower, since $q >> 1$. The latter super-bounce-frequency limit is, in fact, the limit in which the equations of Chew, Goldberger, and Low are applicable.[22] This means that the double adiabatic fluid closures for the pressure tensor elements may be employed with rigor (insofar as $q$ can be considered large). In summary, then, the governing equations for the resulting one-fluid system are the continuity equation, the quasineutrality condition (aka the flux-surface-averaged equation of vorticity), and the double adiabatic pressure equations, viz.,

$$\partial n / \partial t + \vec{\nabla}_\perp \cdot (n\vec{u}_\perp) = 0, \qquad (3)$$

$$\int d\vec{S} \cdot \vec{j}_\perp = 0, \qquad (4)$$

$$d[p_\perp / (nB)] / dt = 0, \qquad (5)$$

$$d[(p_\parallel B^2) / n^3] / dt = 0, \qquad (6)$$

where

$$\vec{u}_\perp = \vec{B} \times \vec{\nabla}\varphi / B^2, \qquad (7)$$

$$\vec{j}_\perp = \vec{B} \times [\vec{\nabla} \cdot (\vec{P}_i + \vec{P}_e) + nMd\vec{u}_\perp / dt] / B^2, \qquad (8)$$

$$d/dt = \partial/\partial t + \vec{u}_\perp \cdot \vec{\nabla}_\perp,$$

where the surface integral in (4) is over a closed magnetic surface. The electron response is adiabatic, as governed by the electron DKE from (1). Thus, $\vec{P}_e$ is isotropic. In addition, $\vec{B}\vec{\nabla} : \vec{P}_e \approx neE_\parallel$, thus consistent with the MHD assumption $E_\parallel << E_\perp$. [Note, in general, parallel stresses may be evaluated as
$\vec{B}\vec{\nabla} : \vec{P} = \vec{B} \cdot \vec{\nabla}p_\perp + B\vec{B} \cdot \vec{\nabla}((p_\parallel - p_\perp) / B) = \vec{B} \cdot \vec{\nabla}p_\parallel - ((p_\parallel - p_\perp)/B)\vec{B} \cdot \vec{\nabla}B$.] Further, in Eqs (3) – (6), we have neglected the variable $u_\parallel$, consistent with the mode frequency being supersonic. We assume that the magnetic field is known and given as
$\vec{B} = RB_T\vec{\nabla}\zeta + \vec{\nabla}\zeta \times \vec{\nabla}\psi$.

The above constitutes a closed system of (nonlinear) equations for the variables $\{n, \varphi, p_\parallel, p_\perp\}$. To describe GAMs, we begin with an equilibrium in which $n = n(\psi)$ and the pressures are isotropic, $p = p(\psi)$, and then consider small perturbations about this equilibrium. We allow unequal electron and ion temperatures in their maxwellians. In this case, the continuity equation becomes

$$\partial (\tilde{n}/n)/\partial t - 2\vec{\tilde{u}}_\perp \cdot \vec{\nabla} B / B = 0 \tag{9}$$

where $\vec{\tilde{u}}_\perp = \tilde{\varphi}'(\psi)\vec{B} \times \vec{\nabla}\psi / B^2$. It is convenient to introduce $G \equiv B^{-3}\vec{B} \times \vec{\nabla}\psi \cdot \vec{\nabla} B$. Then, $\vec{\tilde{u}}_\perp \cdot \vec{\nabla} B / B \equiv G\tilde{\varphi}'$. Next, the linearized quasineutrality equation (4), after some manipulation, becomes

$$< nM |\vec{\nabla}\psi|^2 / B^2 > \partial \tilde{\varphi}'/\partial t = - < B^{-2}\vec{B} \times \vec{\nabla}\psi \cdot [\vec{\nabla}(\tilde{p}_\perp + \tilde{p}_e) + (\tilde{p}_\parallel - \tilde{p}_\perp)\hat{b} \cdot \vec{\nabla}\hat{b}] >, \tag{10}$$

where $\tilde{p}_e = \tilde{n} T_e$, $<f> \equiv \int (dS / |\vec{\nabla}\psi|) f$. Adiabatic response for the electrons has been assumed; $T_e$ is the electron temperature of an equilibrium maxwellian plasma. The linearized double adiabatic equations (5) and (6) become

$$\partial (\tilde{p}_\parallel / p - 3\tilde{n}/n)/\partial t + 2\vec{\tilde{u}}_\perp \cdot \vec{\nabla} B / B = 0, \tag{11}$$

$$\partial (\tilde{p}_\perp / p - \tilde{n}/n)/\partial t - \vec{\tilde{u}}_\perp \cdot \vec{\nabla} B / B = 0. \tag{12}$$

Equations (9) – (12) now constitute a closed set for $\{\tilde{n}, \tilde{\varphi}, \tilde{p}_\parallel, \tilde{p}_\perp\}$, except that $\tilde{\varphi}$ is coupled to weighted averages of $\tilde{p}$. We proceed as follows: from (11) and (12), we may solve for $\tilde{p}_\perp$ and $\tilde{p}_\parallel$ in terms of $\tilde{n}$, using (9). This yields

$$\tilde{p}_\parallel / p = 2\tilde{n}/n, \quad \tilde{p}_\perp / p = (3/2)\tilde{n}/n. \tag{13}$$

We substitute these into Eq (10). We then use the low equilibrium equation $\vec{\nabla}_\perp B^2 = 2B^2 \hat{b} \cdot \vec{\nabla}\hat{b}$ in the last term on the RHS of (10), and perform an integration by parts on the 1st term on the RHS. Equation (10) then becomes

$$nM < |\vec{\nabla}\psi|^2 / B^2 > \partial \tilde{\varphi}'/\partial t = - < G(\tilde{p}_\perp + \tilde{p}_\parallel + 2\tilde{p}_e) >, \tag{14a}$$

$$(\tilde{p}_\perp + \tilde{p}_\parallel)/p = (7/2)\tilde{n}/n, \quad \tilde{p}_e/p = (\tilde{n}/n)(T_e/T_i), \tag{14b}$$

where we have used Eqs (13). Finally, we operate on Eq (9) with the operator $<G.....>$. The resulting equation is

$$(\partial/\partial t)<G(\tilde{n}/n)>=2\tilde{\varphi}'<G^2>. \quad (15)$$

Equations (14) and (15) are now closed. A wave equation for the GAM is easily derived and the frequency of oscillation can be found to be

$$\omega^2 = 2(T_i/M)(7/4+T_e/T_i)<2G^2>/<(|\vec{\nabla}\psi|/B)^2>. \quad (16)$$

Henceforth, we define this to be the GAM frequency, $\omega_{GAM}^2$. Specializing to circular flux surfaces in large aspect ratio, we get

$$\omega^2 = 2(7/4+\tau)T_i/(MR^2) \quad (17)$$

where $\tau \equiv T_e/T_i$. This is in agreement with the real frequency found by Sugama and Watanabe from collisionless gyro-kinetic theory. These authors have also calculated the correction from the sonic effects. Such corrections could presumably be obtained from our CGL theory by taking our calculation to the next order. The Landau damping would, of course, have to be calculated from the DKE. For this paper, we are concerned more with the external driving and the physics of magnetic pumping damping, both examined in the sections following.

**Externally Driven GAMs**

We now revisit the foregoing mode calculation for the GAM and consider external drivers to excite this mode. In particular, we examine two possible drivers: an axisymmetric current coil system that would be capable of shifting the entire set of tokamak flux surfaces in periodic up-down displacements; and an external heating system that would periodically modulate the perpendicular ion pressure, $p_\perp$, at some localized region of plasma (for example, heating ions via ion-cyclotron resonance). To model the flux-surface displacements, we may imagine that the ExB drift in the fluid equations includes an added piece so that

$$\tilde{\vec{u}}_\perp \rightarrow \tilde{\vec{u}}_\perp + \hat{z}d\tilde{\xi}(t)/dt, \quad (18)$$

where $\xi = \xi(t)$. This corresponds to up-down displacements of flux surfaces that are rigid and axisymmetric. For the perpendicular ion heating, we model the heating function by adding a term $H(x,t)$ to the $p_\perp$ equation, *viz.*,

$$\partial \tilde{p}_\perp/\partial t - 3p\tilde{\vec{u}}_\perp \cdot \vec{\nabla}B/B = \tilde{H}(\vec{x},t). \quad (19)$$

Note that $\tilde{H}(x,t)$ need not be axisymmetric. This assumption will be checked later on but it is useful to review the nature of GAMs and zonal flows here. For our purposes in this paper, we assume that the GAM frequency is lower than the shear Alfven frequency, ie, $\omega \ll V_A/qR$. This allows the electrostatic assumption, $\vec{E} \approx -\vec{\nabla}\varphi$. In addition, the electron communication along the lines is very rapid, ie, $\omega \ll v_e/qR$. This leads to adiabatic electron response and, by extension, to $E_\parallel \ll E_\perp$. Consequently, the electric potential is very nearly constant on each flux surface. Since ExB drifts are dominant, and the magnetic field is very nearly constant in time, this means that the dominant flows lie on the flux surface and the "frozen-in" flux surface motions must be consistent with the magnetic geometry. For tokamaks, since the field is axisymmetric and the electric potential is a flux surface, the dominant cross-field plasma flows are also axisymmetric, that is to say these are zonal flows. To be sure, there can be non-axisymmetric parallel flows, $u_\parallel$. Associated with these parallel flows, there may be non-axisymmetric density or pressure fluctuations. However, if the object is to drive zonal flows, one may use non-axisymmetric sources, as the *m=1, n=0* harmonic piece of this source may couple to zonal flows, even as the other harmonics may excite parallel non-axisymmetric sound waves. All this explains why our set of equations above includes non-axisymmetric $\{n, p_\parallel, p_\perp\}$ but axisymmetric $\varphi$.

We consider the two external source terms mentioned above separately, beginning with the ion heating. We insert $\tilde{H}(x,t)$ as in (19) above and let $\tilde{H}(x,t) \to \tilde{H}(x)\exp(-i\omega t)$. We also let $\partial/\partial t \to -i\omega$ in all the linearized equations. It is then straightforward to solve for the amplitude of the density oscillations associated with the driven GAM. We find

$$(1-\omega^2/\omega_{GAM}^2)\tilde{\varphi}' = -<G\tilde{H}/(nT_i)>/[2<2G^2>(7/4+T_e/T_i)] \qquad (20)$$

where $|\tilde{\vec{u}}_\perp| = |\tilde{\varphi}'||\vec{\nabla}\psi/B|$. Note that $\tilde{\varphi}'$ has dimensions of frequency. Note also that $\tilde{H}(x,t)$ must have the same symmetry as $G$ (up-down antisymmetric) to have effective forcing.

To study the resonance arising from axisymmetric flux surface displacements, we now start with equations (9)-(12) and employ the substitution (18). This corresponds to a rigid upward displacement of flux surfaces. All the convected scalars $\{n, p_\parallel, p_\perp\}$ with equilibrium profiles would then undergo the replacement, in their respective equations, $\partial_t \tilde{f} \to \partial_t \tilde{f} + \partial_t \tilde{\xi}\, \partial_z f$. Thus, Eq (9) becomes

$$\partial_t(\tilde{n}+\tilde{\xi}\,\partial_z n) - 2n\tilde{\vec{u}}_\perp \cdot \vec{\nabla}B/B = 0, \qquad (21)$$

and so on. Manipulating the equations as previously, we get the counterparts of Eqs (13), modified by the $\tilde{\xi}$ shifts. This leads to the expression for $\tilde{p}_\perp + \tilde{p}_\parallel + 2\tilde{p}_e$

$$(\tilde{p}_\perp + \tilde{p}_\parallel + 2\tilde{p}_e)/p = 2(7/4+\tau)(\tilde{n}+\tilde{\xi}\,\partial_z n)/n - 2\tilde{\xi}\,(\partial_z p/p + \tau\partial_z n/n). \qquad (22)$$

Substituting this expression into the RHS of (14a) results in an equation that couples $\tilde{\varphi}'$ to a weighted average of $\tilde{n} + \tilde{\xi}\partial_z n$ and, in addition, to an inhomogeneous term proportional to $\tilde{\xi}$. The system of inhomogeneous equations is now closed by taking the $<G.....>$ average of (21). We find an inhomogeneous equation for the driven amplitude as the mode approaches resonance, as follows:

$$nM <|\vec{\nabla}\psi|^2/B^2 > \partial_t^2\tilde{\varphi}' = -2p[7/4+\tau]<2G^2>\tilde{\varphi}' + 2p\partial_t\tilde{\xi}<G(\partial_z p/p + \tau\partial_z n/n)>.$$

Note that the driver term is nonzero since the z-derivative of $p$ and the magnetic term $G$ may have the same up-down symmetry. Assuming solutions of the form $A\exp(-i\omega t)$, where $A$ is the amplitude of the potential, we get the driven amplitude

$$|1-\omega^2/\omega_{GAM}^2\|\tilde{\varphi}'| = \omega|\tilde{\xi}<G(\partial_z p/p + \tau\partial_z n/n)>|/<2G^2(7/4+\tau)>. \qquad (23)$$

**Damping from magnetic pumping**

The GAM oscillation involves motion of toroidal flux tubes back and forth between regions of differing B. As a consequence, energy is "pumped" into $p_\perp$ during a compression. If in the course of this compression there is a collision, some of the $p_\perp$ transfers to $p_\parallel$. The process is random, but, in a random walk, a portion of the $p_\perp$ energy pumped is not fully recovered to restore the flux tube back to its original position, resulting in a damping of the oscillation.[17-21] This is the well-known damping from magnetic pumping. For the subsonic, zero frequency, poloidal rotation, this damping has been calculated in both collisional[17,18] and collisionless regimes.[19] Here, we examine this damping in the collisionless regime for the higher frequency GAMs.

A simple heuristic extension of the CGL equations, introduced earlier,[20,21] can model the tendency to equilibrate $p_\perp$ and $p_\parallel$. These model equations are

$$(d/dt)[p_\perp/(nB)] = -(\nu_{ii}/3)(p_\perp - p_\parallel)/(nB), \qquad (24)$$

$$(d/dt)[p_\parallel B^2/n^3] = (2\nu_{ii}/3)(p_\perp - p_\parallel)B^2/n^3. \qquad (25)$$

The tendency to equilibrate temperature anisotropy in these equations is clearly included. Note that if there is a collision, given a certain pressure anisotropy, it tends to transfer twice as much energy into the parallel temperature than out of the perpendicular temperature, consistent with total kinetic energy conservation and that the perpendicular kinetic energy constitutes two degrees of freedom. We use these model equations in the above normal mode derivation. From the linearization of (24) and (25), we can easily find $p_\perp + p_\parallel$ and we find that that the collisional correction is of the form $\nu_{ii}(p_\perp - p_\parallel)$. To simplify the calculation, since $\nu_{ii}$ is small, we may replace $p_\perp - p_\parallel$ by the lower order expressions (13). Thus, we use the lower order result

$$(\tilde{p}_\perp - \tilde{p}_\parallel)/p \approx -(1/2)\tilde{n}/n \qquad (26)$$

to find the collision-corrected expression for $p_\perp + p_\parallel$ to be

$$(\tilde{p}_\perp + \tilde{p}_\parallel)/p \approx [7/2 - i\nu_{ii}/(6\omega)]\tilde{n}/n. \qquad (27)$$

Substituting this on the RHS of equation (14a), we get a dispersion relation similar to the one previously but with the collision correction. This dispersion is

$$\omega^2 = 2 < 2G^2 > [(7/4 + \tau) - i\nu_{ii}/(12\omega)]. \qquad (28)$$

This yields the relatively weak damping rate of $\nu_{ii}/[24(7/4+\tau)]$. Since Landau damping rates have exponential terms multiplying them, for large safety factor, *q*, these rates could also be small. In this case, collisional damping from magnetic pumping could be considered.

**Summary**

GAMs in tokamaks are super-sonic in the sense that the GAM frequency is higher than the parallel ion bounce frequency (assuming large safety factor *q*). Consequently, these modes are describable, with rigor, by the double-adiabatic closure of the collisionless strongly magnetized fluid equations with anisotropic pressure introduced by Chew, Goldberger, and Low.[1] This approach has the advantage of relative simplicity and allows a ready lowest order nonlinear formulation. We have used this approach in this paper to obtain the homogenous normal mode GAM, and expressions for the mode amplitude near resonance when driven by external sources. We have shown that the real frequency of the mode agrees with that from earlier kinetic theories. We have also used an extension of the double-adiabatic model[20,21] corrected for weak collisions to obtain the collisional damping from "magnetic pumping". The double-adiabatic approach has the drawback that Landau effects cannot be described. However, since the Landau damping rates are attenuated by an exponentially small factor, there could be parameter regimes where the magnetic pumping is important.

Sonic corrections to the GAM real frequency have also been calculated in earlier kinetic theory calculations (in addition to the Landau rates).[11,12] These corrections should be derivable from the CGL equations if these are taken to first-order in the sonic corrections. We have not attempted this in the present paper. Finally, we note that an arbitrary (in three space dimensions) initial condition on the density or pressure perturbation can result in GAMs as well as Landau damped parallel acoustic modes, with the GAM oscillations being on a more rapid time scale. The response to such an initial condition cannot be fully described by only the CGL equations – only the GAM part that couples to the zonal potential is describable by CGL (the *m=1* parts of the perturbation). The parallel acoustic response (that is all the other *m* modes) would have to be described on a slower time scale by the full DKE.   Nonetheless, we note that an external heating source that is not axisymmetric (as well as not up-down symmetric) could resonantly excite GAMs, the non-axisymmetry of the source notwithstanding.

Finally, we note that the results of this paper could be applied, under certain conditions, to non-axisymmetric toroids such as stellarators. [24,25] We observe that the calculation in the paper only makes an assumption of the existence of magnetic surfaces, $\psi$; beyond that, the magnetic field $\vec{B}$ is general, with no assumptions made on axisymmetry.  The key geometric parameter in the calculation, $G \equiv B^{-3}\vec{B}\times\vec{\nabla}\psi\cdot\vec{\nabla}B$, carries over to a non-axisymmetric calculation.   As mentioned, however, certain conditions must apply:  in particular, the use of CGL theory assumes that the mode frequency is super-parallel sonic.  In the case of tokamaks, the latter applies if the parallel connection length, *qR,* is greater than *R*, i.e., *q >> 1.*   In the case of a stellarator, this translates to the condition that the parallel variation in *B* be smaller than the perpendicular variation, i.e, on average, $|\vec{B}\times\vec{\nabla}\psi\cdot\vec{\nabla}B|\gg|\vec{\nabla}\psi||\vec{B}\cdot\vec{\nabla}B|$.   Clearly, if at all, this may only be true in special cases.

## Acknowledgments

We are grateful to Drs. Hallatschek and McKee for sharing with us information on the proposed GAM experiment. Useful discussions with Dr. Hallatschek are gratefully acknowledged.  This work was supported by the US-DOE.